\documentclass{iau_FM}
\usepackage{graphicx}
\usepackage{natbib}

\title[2D-models of early-type fast rotating stars]
{Two-dimensional models of early-type fast rotating stars: the ESTER
project}

\author[Michel Rieutord]   
{Michel Rieutord$^{1,2}$}

\affiliation{$^1$Universit\'e de Toulouse; UPS-OMP; IRAP; Toulouse,
France \\[\affilskip]
$^2$CNRS; IRAP; 14, avenue Edouard Belin, F-31400 Toulouse, France\\
email: {\tt Michel.Rieutord@irap.omp.eu} }

\pubyear{2015}
\jname{Astronomy in Focus, Volume 1}
\editors{Claudia Maraston, ed.}
\begin{document}

\maketitle

\begin{abstract}
In this talk I present the latest results of the ESTER project that
has taken up the challenge of building two dimensional (axisymmetric) models
of stars rotating at any rotation rate. In particular, I focus on
main sequence massive and intermediate mass stars. I show what should
be expected in such stars as far as the differential rotation and the
associated meridional circulation are concerned, notably the emergence of
a Stewartson layer along the tangent cylinder of the core. I also
indicate what may be inferred about the evolution of an intermediate-mass
star at constant angular momentum and how Be stars may form. I
finally give some comparisons between models and observations of the
gravity darkening on some nearby fast rotators as it has been derived
from interferometric observations. In passing, I also discuss how 2D
models can help to recover the fundamental parameters of a star.

\keywords{Stars, rotation}
\end{abstract}

\firstsection 
\section{Introduction}

Rotation has now become an unavoidable parameter of stellar models, but
for most massive or intermediate-mass stars rotation is fast, at least
of a significant fraction of the critical angular velocity.  Current
spherically symmetric models try to cope with this feature of the stars
using various approximations, like for instance the so-called shellular
rotation usually accompanied with a diffusion that is meant to represent
the mixing induced by rotationally generated flows. Such approximations
may be justified in the limit of slow rotation where anisotropies and
associated flows are weak \cite[][]{zahn92}. However, when rotation
is fast, say larger than 50\% of the critical velocity the use of a
spherically symmetric 1D-model is doubtful. This is not only because of
the centrifugal flattening of the star, but also because of the flows
that are induced by the baroclinic torque that naturally appears when the
stable stratification of a radiative region meets rotation. These flows
face the cylindrical symmetry of the Coriolis force and the spheroidal
symmetry of the effective gravity.

The breaking  of spherical symmetry is also to be taken into
account when the early-type star at hands show delta Scuti
type oscillations, like $\alpha$ Aql or $\alpha$ Oph actually do
\cite[][]{monnier_etal10}. The modelling of these seismic data requires
2D models \cite[][]{RLR06}. In the same line,  more than half a dozen of
nearby early-type stars, showing fast rotation, have been observed with interferometers
\cite[e.g.][]{che_etal11}. Here too, 1D models are not usable for data
processing.

Hence, for all these stars 2D models are crucial. Rotation is actually the
main phenomenon that breaks the spherical symmetry on the large scales.
It can be properly accounted for with axisymmetric 2D models.

The challenge of computing 2D models, which unlike previous attempts
include the large-scale flows, has been taken up with the ESTER
project\footnote{\tt http://ester-project.github.io/ester/}. The
first 2D models from this project are giving the structure of an
early-type star including its differential rotation and the associated
meridional circulation \cite[][]{ELR13}. We recall that these flows are
consubstantial to rotating stars and are driven by the baroclinicity of
the radiative envelope \cite[][]{R06}. We show in Fig.~1 the shape of the
differential rotation of a massive star rotating almost critically. Such
results show that baroclinic flows generated by rotation are complex and
far from the one-dimensional shellular rotation imposed in the 1D set-up.
The comparison of 2D models and interferometric data has been particularly
successful \cite[][]{R13,domiciano_etal14}, showing that gravity darkening
can be nicely modeled by a simple law much better than with von Zeipel
rule \cite[][]{ELR11,R15}.

Present ESTER models compute the steady state of an isolated non-magnetic
early-type star. The next step is to implement time-dependence so as to
be able to compute genuine stellar evolution in two-dimensions. The
ultimate evolution of the code will be the shift to 3D models allowing
large-scale magnetic fields and/or tidal distortion to be included.

\begin{figure}[t]
\begin{center}
 \includegraphics[width=0.99\linewidth]{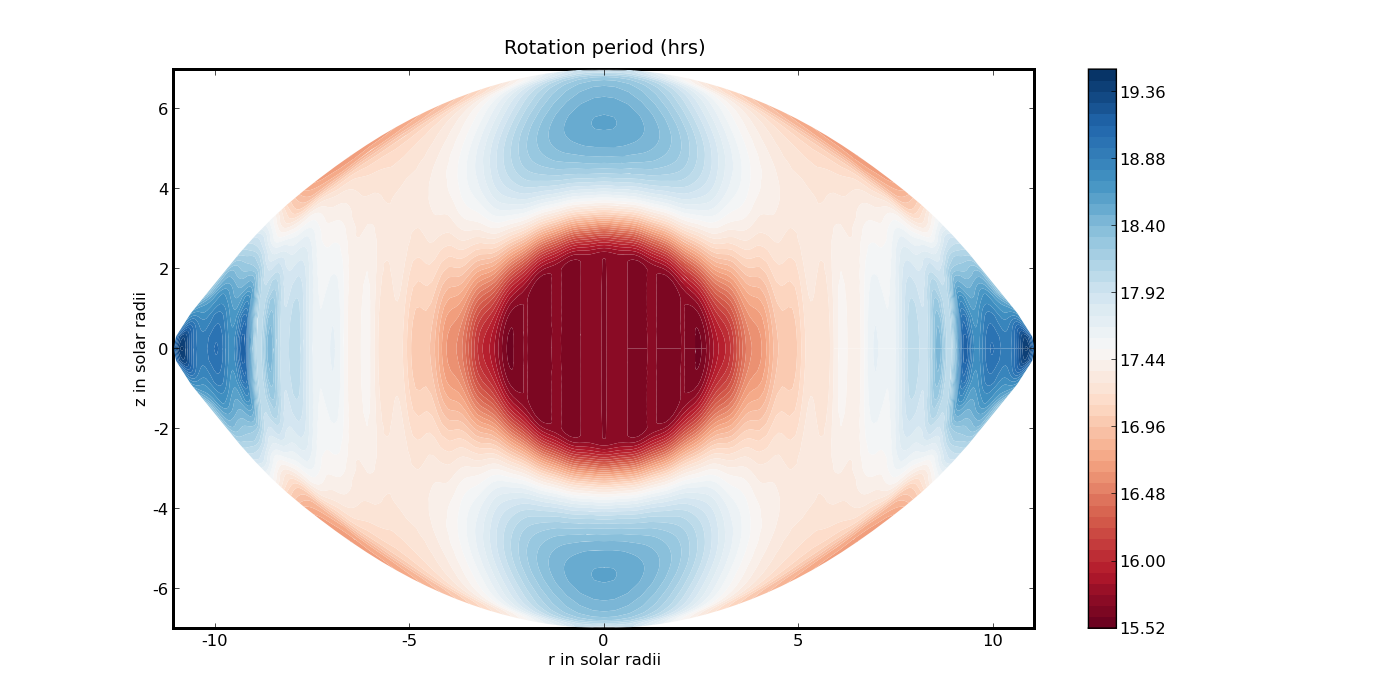}
 \caption{Differential rotation of a 30~M$_\odot$ star rotating at 98\%
of the critical angular velocity. The star is homogeneous chemically
with X=0.7 and Z=0.02.}
   \label{fig1}
\end{center}
\end{figure}


\end{document}